\documentclass[%
  reprint,
  superscriptaddress,
  amsmath,amssymb,
  aps,
  prl,
]{revtex4-2}

\usepackage{graphicx}
\graphicspath{{figures/}}
\usepackage{bm}
\usepackage{mathtools}
\usepackage{xcolor}
\usepackage{cancel}
\usepackage{dcolumn}

\usepackage[colorlinks=true,allcolors=blue]{hyperref}

\def\be{\begin{equation}}
\def\ee{\end{equation}}
\def\bea{\begin{eqnarray}}
\def\eea{\end{eqnarray}}

\begin{document}

\title{Angular Momenta in $4+1$ Vacuum Critical Collapse}

\author{\'Oscar J. C. Dias}
\email{ojcd1r13@soton.ac.uk}
\affiliation{STAG Research Centre and Mathematical Sciences, Highfield Campus, University of Southampton, Southampton SO17 1BJ, UK}
\author{Carsten Gundlach}
\email{C.J.Gundlach@soton.ac.uk}
\affiliation{STAG Research Centre and Mathematical Sciences, Highfield Campus, University of Southampton, Southampton SO17 1BJ, UK}
\author{Jorge E. Santos}
\email{jss55@cam.ac.uk}
\affiliation{Department of Applied Mathematics and Theoretical Physics, University of Cambridge, Wilberforce Road, Cambridge CB3 0WA, UK}
\author{Benson Way}
\email{bway@umd.edu}
\affiliation{Maryland Center for Fundamental Physics, University of Maryland, College Park, MD 20742, USA}

\date{\today}

\begin{abstract}
We investigate the critical collapse of gravitational waves that carry angular momenta.  Working in five-dimensional vacuum Einstein gravity under an SU$(2)\times\mathbb Z_4$ symmetry ansatz, we numerically evolve several families of initial data near the threshold of black-hole formation.  Near-critical data approaches the discretely self-similar Bizo\'n-Chmaj-Schmidt solution, which is irrotational.  We find that the near-threshold mass scaling has a universal critical exponent, but the angular-momentum scaling has one of two critical exponents, depending on the family of initial data.  The angular-momentum critical exponents are therefore not universal under this symmetry ansatz.  Our results agree with predictions from perturbation theory about the critical solution.  
\end{abstract}

\maketitle
{\bf Introduction.} Any generic configuration of a spherically symmetric massless scalar field in flat space eventually collapses into a black hole or disperses away.  Initial data can be fine-tuned between these two outcomes, where evolution then approaches a universal, discretely self-similar solution \cite{Choptuik:1992jv,Gundlach:1995kd,Gundlach:1996eg}. For initial data parametrised by $p$ near the threshold value $p_\star$, the black hole mass scales as $M\propto(p-p_\star)^\gamma$, for critical exponent $\gamma$.  The mass also shows a periodic modulation with period $T$ as it approaches criticality.  Both $\gamma$ and $T$ are universal in that they do not depend on the family of initial data.  Rather, these are set by the critical solution, where $\gamma$ is determined by the critical solution's unstable mode, and $T$ depends on the echoing period that defines the discrete self-similarity.  Critical behaviour was soon observed in vacuum gravity, first with axisymmetric gravitational-waves in four dimensions \cite{Abrahams:1993Vacuum},\footnote{The interpretation of these results as evidence for a universal discretely self-similar critical solution has been called into question by recent high-resolution studies. Independent numerical codes find only approximate self-similarity for some families of initial data and, so far, no evidence for a unique universal critical solution in four-dimensional axisymmetric vacuum gravity \cite{Baumgarte2023,BaumgarteGundlachHilditch2026,Adhikari2026}.} and later in five-dimensions \cite{Bizon:2005Critical}.

The effect of angular momentum on critical collapse is less well understood. Perturbation theory predicts that close to a non-rotating critical solution, the black hole angular momentum scales as $J\propto(p-p_\star)^\mu$, with $\mu$ determined by the least-damped mode that couples to $J$ \cite{Gundlach:1998Angular,Gundlach:2001ey}. If no mode carries angular momentum at linear order, $J$ can instead arise at second order, as happens for a massless scalar field \cite{Garfinkle:1999Angular}. These ideas have subsequently been developed and tested in a variety of matter models with angular momentum, including rotating radiation and perfect fluids \cite{Baumgarte:2016Rotating,GundlachBaumgarte2016,GundlachBaumgarte2018},
and scalar fields \cite{Choptuik:2004Angular,Olabarrieta2007,JalmuznaGundlach2017}. However, critical collapse with angular momentum in the most universal sector of pure vacuum gravity has, to our knowledge, not been studied; the twist-free axisymmetric settings exclude it by construction.

We investigate this phenomenon in five spacetime dimensions, where an SU$(2)$ symmetry on the three-spheres reduces the Einstein equation to a $(1+1)$-dimensional system. Using numerical time evolution within a non-rotating sector of this symmetry class, Bizo\'n, Chmaj and Schmidt (BCS) found a universal discretely self-similar critical solution with a single unstable mode \cite{Bizon:2005Critical}. This symmetry reduction can be extended to include angular momentum. The SU$(2)\times\mathbb{Z}_4$ ansatz of Ref.~\cite{Crump:2026kgu} (partially inspired by Ref.~\cite{Ishii:2018oms}) supports the BCS critical sector while also allowing for gravitational degrees of freedom that carry angular momentum.  The dynamical equations remain $(1+1)$-dimensional. 

In this Letter we numerically evolve several families of vacuum initial data carrying angular momentum and tune them to the threshold of black-hole formation. We find that all families near threshold approach the BCS critical solution, but the scaling of angular momentum depends on the family of initial data, and is therefore not universal within the symmetry restrictions.  

We compare our results with perturbation theory. We numerically construct the BCS critical solution directly under a discretely-self-similar ansatz and calculate its linear perturbations. To our knowledge, this construction in this system is also new \cite{Dias:2026Companion}.  Our perturbative and time-evolution results agree quantitatively.  

Perturbation theory explains the non-universal scaling behavior of angular momentum.  There are two sets of perturbations that respect the imposed symmetries: one where the rotation axis is aligned along a principal axis of the critical solution, and another where the axes are orthogonal.  Both perturbations generate angular momentum, though the aligned orientation only at second order.   Within the ansatz, symmetry forbids a smooth interpolation between the two perturbations, so each has its own distinct critical scaling of angular momentum.  We expect the orthogonal (tilted) configuration to be more generic under relaxed symmetry assumptions because it is the dominant mode.

{\bf Formulation.} Consider five-dimensional vacuum spacetimes, $R_{\mu\nu}=0$, with an SU$(2)$ isometry on three-dimensional orbits. Let $\sigma^i$ be the left-invariant one-forms on SU$(2)$, with ${\rm d}\sigma^i=\tfrac12\epsilon^{i}{}_{jk}\sigma^j\wedge\sigma^k$.  Take the ansatz
\begin{multline}
  {\rm d}s^2=g_{ab}\,{\rm d}y^a {\rm d}y^b+\frac{r^2}{4}e^{B}\Big[\sqrt{1+|\Phi|^2}\,\sigma_+ \sigma_-
\\
+\frac{\overline{\Phi}}{2}\sigma_+^2+\frac{\Phi}{2} \sigma_-^2+e^{-3B}\left(\sigma^3+2 \mathcal{A}\right)^2\Big],
  \label{eq:ansatz}
\end{multline}%
where $\sigma_{\pm}\equiv\sigma^1\pm {\rm i}\sigma^2$ and all fields depend on the coordinates $y^a=(t,r)$. 
Here $r$ is the areal radius, $B$ describes the biaxial squashing of the three-sphere, while the complex scalar field $\Phi=X+{\rm i}Y$ deforms the sphere in the $\sigma^1-\sigma^2$ plane. For $\Phi=\mathcal A=0$, Eq.~(\ref{eq:ansatz}) reduces to the biaxial Bianchi IX ansatz of BCS~\cite{Bizon:2005Critical}, whereas for $\Phi=0$, $\mathcal A\neq0$, it contains equal-angular-momenta Myers-Perry black holes \cite{Myers:1986un,Gibbons:2004uw,Gibbons:2004js,Kunduri:2006qa,Dias:2010eu}.  $g_{ab}$ contains residual gauge symmetry which must be fixed for numerical solutions.

\begin{figure}[t]
  \centering
  \includegraphics[width=\columnwidth]{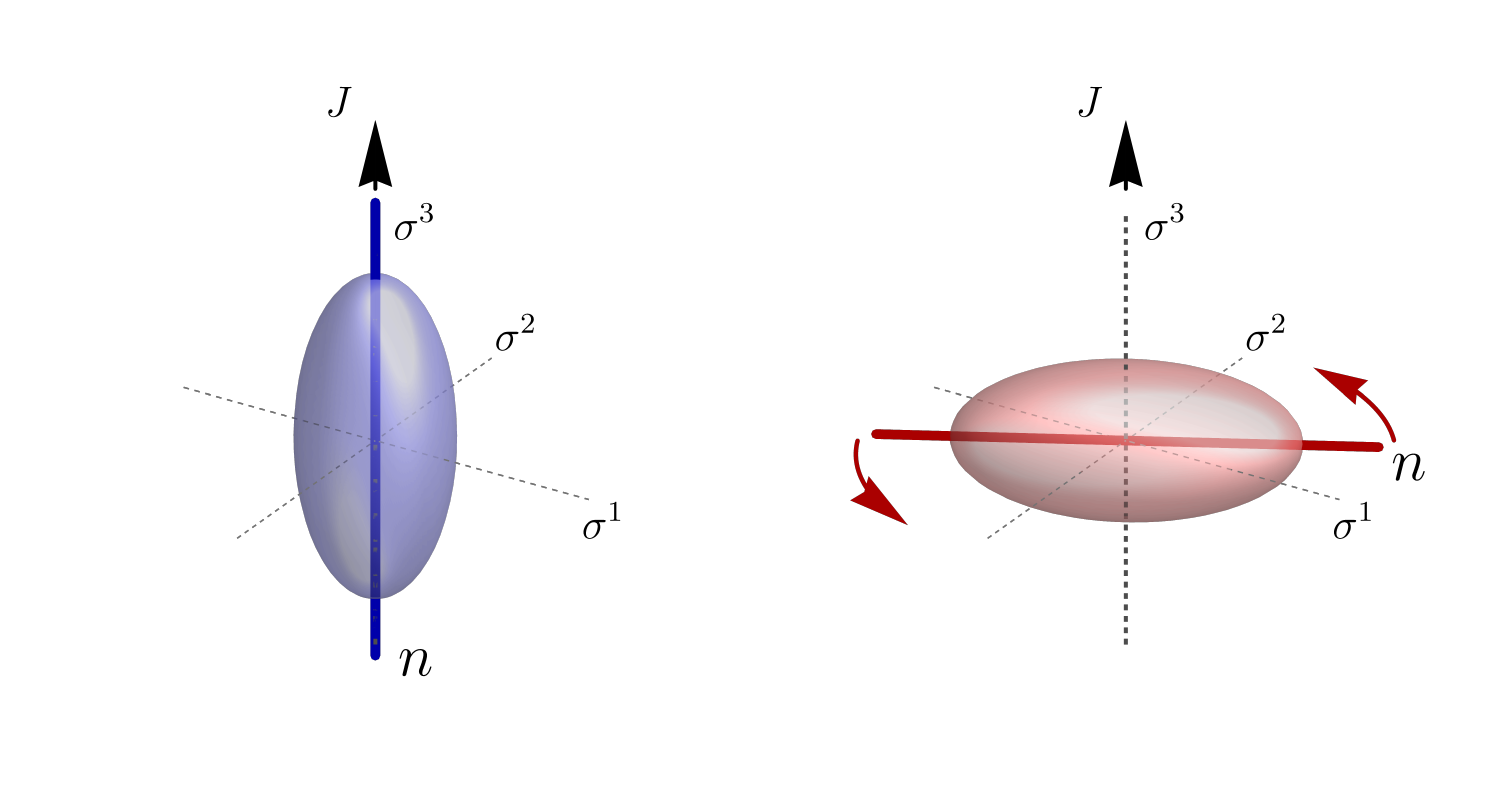}
  \caption{The two embeddings of the biaxial BCS solution in the SU$(2)\times\mathbb Z_4$ ansatz. The rotation axis $J$ is along $\sigma^3$, which is always an eigendirection of the orbit metric, so the symmetry axis $n$ of the critical solution is either aligned with it, $n\parallel\sigma^3$ (left panel), or orthogonal to it, $n\perp\sigma^3$ (right panel).  Arrows 
  in the right panel indicate the rotation of $n$ about $\sigma^3$, the tilt mode that carries the angular momentum.}
  \label{fig:orientation}
\end{figure}

Though the SU$(2)\times\mathbb Z_4$ ansatz admits the BCS biaxial configuration, it does so in two inequivalent orientations relative to the rotation axis $\sigma^3$, as we depict in Fig.~\ref{fig:orientation}. The logarithms of the eigenvalues 
of the orbit metric are
\begin{equation}
  B+\operatorname{arcsinh}|\Phi|,\qquad
  B-\operatorname{arcsinh}|\Phi|,\qquad
  -2B .
\end{equation}
Biaxiality requires two of these to be equal, which occurs for $\Phi=0$, when the symmetry axis is aligned with the rotation axis, or for $\operatorname{arcsinh}|\Phi|=3|B|$, when the symmetry axis and rotation axis are orthogonal.

Nonzero $\Phi$ breaks the U$(1)$ symmetry generated by right translations along $\sigma^3$ to $\mathbb Z_4$, giving the SU$(2)\times\mathbb Z_4$ symmetry class introduced in Ref.~\cite{Crump:2026kgu}. The vacuum Einstein equation reduces to a $(1+1)$-dimensional system. 

The complex field $\Phi$ is charged under the reduced gauge field $\mathcal A$, and its associated current carries angular momentum. Therefore, define $F={\rm d}\mathcal A$ and $D_a\Phi=(\partial_a+4{\rm i}\mathcal A_a)\Phi$.  In $G_5=1$ units, the angular momentum enclosed by an orbit of areal radius $r$ may be written as
\begin{equation}
  J(r)
  =-\frac{\pi}{16}\,r^5e^{-2B}\star F
  =\frac{\pi}{4}\int_0^r \tilde{r}^3\,
\mathrm{Im}\left(\bar\Phi\,\Pi_\Phi\right)\,{\rm d}\tilde{r},
  \label{eq:J}
\end{equation}
where $\Pi_\Phi\equiv \hat n^aD_a\Phi/|Dr|=\Pi_X+{\rm i}\Pi_Y$. Here $\hat n^a$ is the future-directed unit normal to the radial slice and $|Dr|$ is the norm of the gradient of $r$ within the slice. Similarly define $\Pi_B\equiv \hat n^a\partial_aB/|Dr|$. Thus, angular momentum is carried entirely by the $\Phi$ sector. 
When $\Phi=0$, as for the final black hole, $J(r)$ coincides with the Komar angular momentum.

{\bf Methods.} We choose the initial data
\begin{multline}
  B=2c\,\varepsilon f,\quad
  \Phi={\rm i}\,\varepsilon f,\quad
  \Pi_X=5\,\varepsilon f,
  \\
  \Pi_B=\Pi_Y=0,\quad
  f(r)=\frac{r^2e^{-r^2/2}}{(1+r^2)^{7/4}},
  \label{eq:ID}
\end{multline}
where $\varepsilon$ and  $c$ are two independent parameters. At fixed $c$, we bisect in the amplitude $\varepsilon$ and then evolve sequences of supercritical data with offsets $\delta\equiv\varepsilon-\varepsilon_\star$ between $10^{-8}$ and $4\times10^{-5}$: 39 values at $c=0$ and 16 at $c=1$.

We use two independent numerical implementations and gauge choices: an $hp$-adaptive nodal discontinuous-Galerkin code (DG) in maximal-areal gauge (already used to study gravitational collapse in AdS \cite{Choptuik:2017cyd}, and a fourth-order finite-difference code (FD) with mesh refinement in polar-areal gauge. The two codes give consistent threshold amplitudes $\varepsilon_\star$ and agree on the angular-momentum critical exponents $\mu$ and periods $T_J$; the mass critical exponent $\gamma$ and its period $T$ are measured with the discontinuous-Galerkin code.  Further details of the gauge choices, equations of motion, and numerical validation can be found in the Supplemental Material. 

In maximal-areal gauge (maximal slicing with areal radius, as in~\cite{Choptuik:1999gh}), we locate black holes as the outermost marginally trapped orbit, with areal radius $r_+$, and set $J_{\rm BH}=J(r_+)$ from Eq.~(\ref{eq:J}), averaged once the hole has settled. In the alternative polar-areal gauge, a trapped orbit creates a coordinate singularity; there we read $r_+$ and $J$ at the minimum of the redshift factor $A$ once $\max(1-A)>0.9$, a proxy adequate for $J_{\rm BH}$ but not for the mass. The mass comes from the equal-angular-momenta Myers-Perry relation (with $G_5=1$):
\begin{equation}
  M_{\rm BH}=\frac{3\pi}{8}
  \left(r_+^3+\frac{16J_{\rm BH}^2}{\pi^2r_+^3}\right)^{2/3}.
  \label{eq:MMP}
\end{equation}

Finally, for comparisons with perturbation theory, the BCS solution is constructed as an exactly discretely self-similar solution of the biaxial equations.  We do so by solving a periodic boundary-value problem with unspecified echoing period. Linear perturbations about this background can then be solved.  We use Newton iteration for both background and perturbations \cite{Dias:2026Companion}.

An output of this calculation includes the echoing period $\Delta$ and perturbative modes $\lambda$, which can be compared against the periods and critical exponents of the time evolution data.  More precisely, let $\tau$ be the logarithmic time of the discrete self-similar solution, so that one echo is $\tau\to\tau+\Delta$ and perturbations evolve as $e^{\lambda\tau}$. An evolution with offset $\delta$ departs from the critical solution when the dominant unstable mode, with amplitude $\propto\delta\,e^{\lambda_u\tau}$ (with $\lambda_u=T/\Delta$), becomes of order one. This occurs at $e^{-\tau_\star}\propto\delta^{1/\lambda_u}$, which sets the length scale $e^{-\tau_\star}$ of the resulting black hole. By dimensional analysis, mass scales as $e^{-2\tau_\star}$, while angular momentum scales as $e^{-3\tau_\star}$.

{\bf Results.}  We concentrate on $c=0$, whose data start with no squashing ($B=0$), and on $c=1$.  Near threshold, the evolutions exhibit discrete echoing at the centre before dispersing or forming a black hole. 
\begin{figure}[t]
  \centering
  \includegraphics[width=0.49\columnwidth]{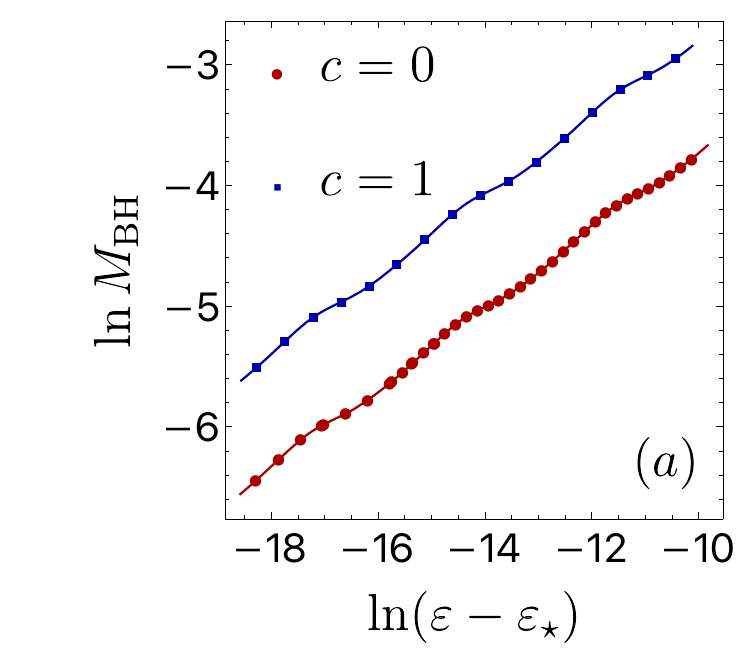}\hfill
  \includegraphics[width=0.49\columnwidth]{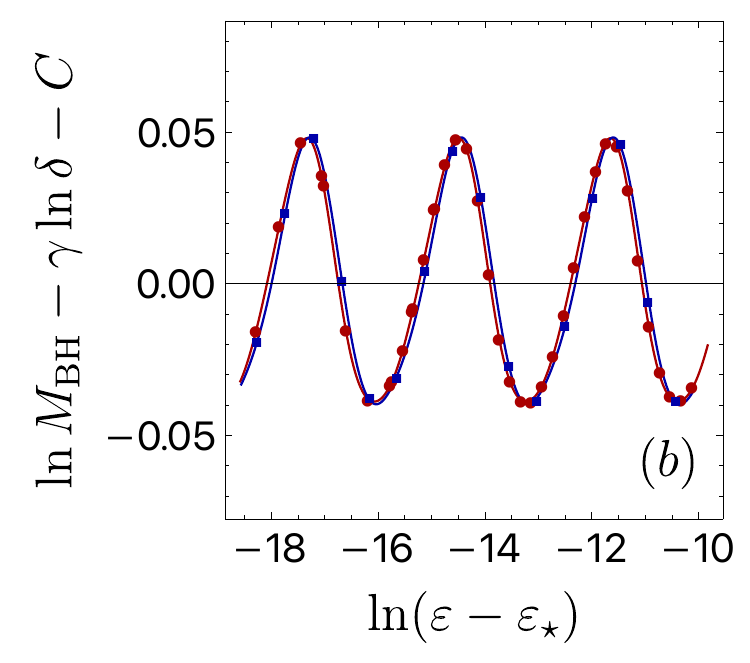}\\[4pt]
  \includegraphics[width=0.49\columnwidth]{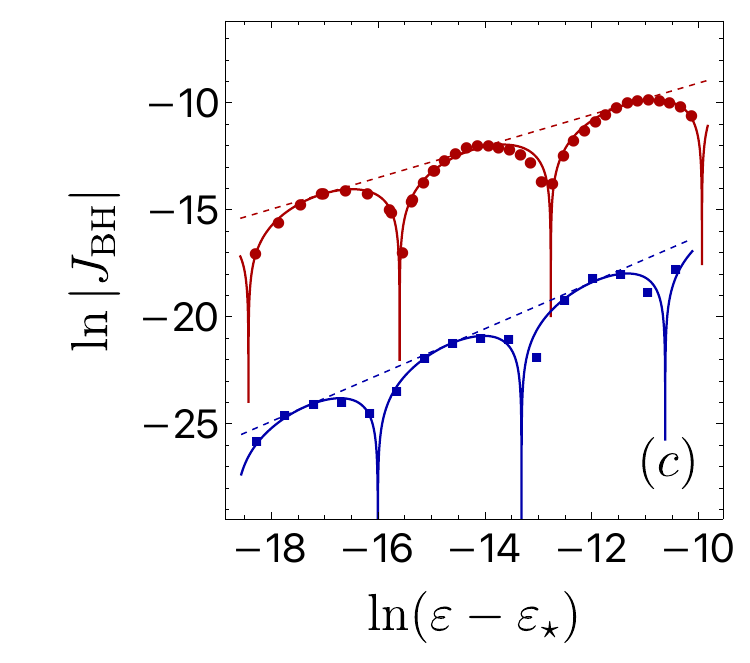}\hfill
  \includegraphics[width=0.49\columnwidth]{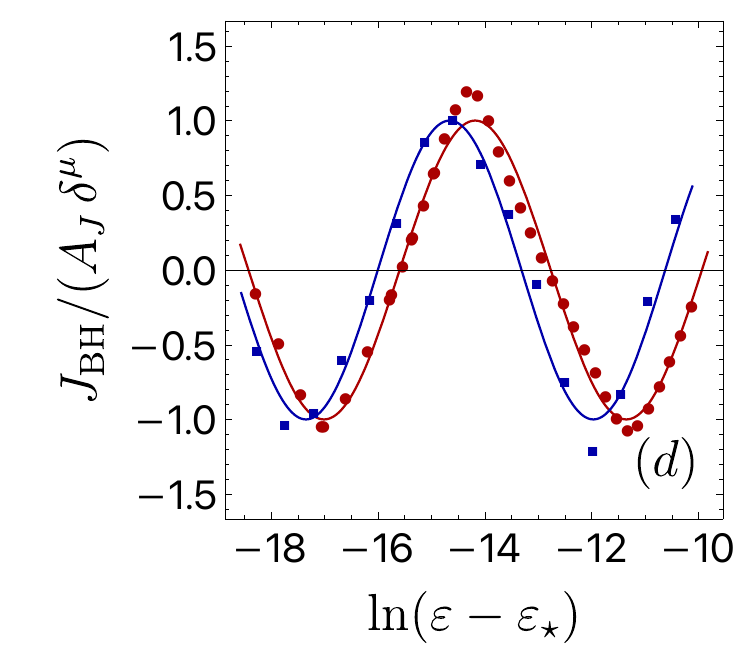}
  \caption{Near-threshold scaling of the initial data (\ref{eq:ID}) with $c=0$ (red circles, tilted) and $c=1$ (blue squares, aligned), against $\ln\delta\equiv\ln(\varepsilon-\varepsilon_\star)$. 
  (a) Black-hole mass (\ref{eq:MMP}), with fits to (\ref{eq:Mfit}) over all points (Table~\ref{tab:exponents} quotes the most fine-tuned window). (b) The same data with the power law removed, and the fitted periodic fine structure $\psi$; both families show the same period $T$. (c) Angular momentum, with fits to (\ref{eq:Jfit}) (solid) and their envelopes (dashed). (d) Angular momentum divided by its envelope, with the fit with period $T_J$. $J_{\rm BH}$ changes sign once per period $T$ of the mass modulation in (b), i.e. $T_J\simeq2T$.}
  \label{fig:scaling}
\end{figure}
Figure~\ref{fig:scaling}(a) shows the black-hole mass $M_{\rm BH}$, and Fig.~\ref{fig:scaling}(b) the same data with the power law removed. Both families ($c=0,1$) are well-fitted by
\be
  \ln M_{\rm BH}=\gamma\ln\delta+C+\psi(\ln\delta),
  \label{eq:Mfit}
\ee
with $\psi$ periodic with period $T$. Representing $\psi$ by a truncated Fourier series 
(three harmonics for $c=0$ and two for $c=1$), both families of data give the same critical exponent $\gamma$ and period $T$ to numerical uncertainty (Table~\ref{tab:exponents}).

This agreement shows that both approach the same critical solution. The critical exponent and period agree with the BCS results \cite{Bizon:2005Critical}, see also \cite{Veronese:2022fnv}.  We can further compare the time evolution results with the direct construction of the BCS solution.  There, we find the echoing period and a single single growing Floquet mode:
\begin{equation}
    \Delta=0.46717031\;,\qquad \lambda_u=6.086(1)\;.
\end{equation}
Because mass has dimensions of length squared in five dimensions, $\gamma=2/\lambda_u$, and the periodic modulation $\psi$ has period $T=\lambda_u\Delta$, corresponding to one echo. Both $\gamma$ and $T$ agree with time evolution data to $10^{-3}$. The combination $\gamma T/2$, which equals $\Delta$ (independent of $\lambda_u$) also agrees to $10^{-3}$. (See Table~\ref{tab:exponents}.)  We conclude that both families of initial data approach the BCS critical solution.

\begin{table}[t]
  \begin{ruledtabular}
  \begin{tabular}{llccc}
   Orientation & & Floquet & DG & FD \\
    \hline
    Tilted ($c=0$)  & $\gamma$ & 0.3286(1)  & 0.32885(6)  & -- \\
            & $T$      & 2.8432(5)  & 2.8408(17)  & -- \\
            & $\mu$    & 0.743(2)   & 0.737(9)    & 0.740(7) \\
            & $T_J$    & 5.686(1)   & 5.66(5)     & 5.72(2) \\
    \hline
    Aligned ($c=1$) & $\gamma$ & 0.3286(1)  & 0.32877(19) & -- \\
            & $T$      & 2.8432(5)  & 2.8415(61)  & -- \\
            & $\mu$    & 1.088      & 1.089(25)   & 1.088(6) \\
            & $T_J$    & 5.686(1)   & 5.62(12)$^{\rm a}$ & 5.61(7)
  \end{tabular}
  \end{ruledtabular}
  \raggedright
  \footnotesize $^{\rm a}$Omitting the four least fine-tuned points ($\ln\delta>-12$); all 16 points give $5.38(9)$.
  \caption{Critical exponents and periods. Floquet: predictions from the 
  DSS 
  solution and its perturbations: see (\ref{eq:mutilt}) and (\ref{eq:mualigned}).  
  DG and FD: the discontinuous-Galerkin and finite-difference time evolutions. 
  The tilted and aligned rows are for the family (\ref{eq:ID}) with $c=0$ and $c=1$, respectively.
  } 
  \label{tab:exponents}
\end{table}

Nevertheless, the two families approach the BCS solution in different orientations. Recall that when $\Phi=0$, the symmetry axis $n$ of the Biaxial ansatz is parallel to the rotation axis $\sigma^3$ (i.e. $n\parallel\sigma^3$).  When instead $\operatorname{arcsinh}|\Phi|=3|B|$, these axes are othogonal, ($n\perp\sigma^3$).  The orientation can therefore be distinguished by analysing $\Phi$ and $B$.  
For this purpose, identify $s_1\le s_2\le s_3$ with the 
eigenvalues of the orbit metric $\{e^{B \pm \operatorname{arcsinh}|\Phi|},e^{-2B}\}$ and define the anisotropy $(s_3-s_1)/s_2$.

For $c=0$ at $r=0.01$, the ratio $\operatorname{arcsinh}|\Phi|/|B|$ remains close to $3$ throughout the echoes, the anisotropy grows to $0.6$, and the two nearly equal stretches agree to $2\times10^{-4}$. The $c=0$ case therefore approaches the tilted orientation ($n\perp\sigma^3$).  

For $c=1$, the ratio $|\Phi|/|B|\le2\times10^{-3} \simeq 0$ at $r=0.01$; and at $r=0.1$, it decreases from echo to echo, so we conclude that this data approaches the aligned orientation ($n\parallel\sigma^3$).  

The orientation depends strongly on the choice of initial data.  A survey with the finite-difference code finds the tilted orientation not only at $c=0$ but also at $c=-0.3$ and $0.3$, and the aligned one at $c=-1$ and $3$, in addition to $c=1$ (Supplemental Material).

We now investigate the angular-momentum scaling [Fig.~\ref{fig:scaling}(c,d)]. For both families we find
\be
  J_{\rm BH}\simeq A_J\,\delta^{\mu}\cos\left(\frac{2\pi}{T_J}\,\ln\delta+\varphi\right).
  \label{eq:Jfit}
\ee
Interestingly, in both $c=0$ and $c=1$ cases, $T_J\simeq2T$, so $J_{\rm BH}$ changes sign once per echo. 
Results for $\mu$ and $T_J$ are summarised in Table~\ref{tab:exponents}. We note that $T_J$ agrees well between $c=0$ and $c=1$ families, but $\mu$ does not.  

We now compare our dynamical results with perturbation theory about the BCS critical solution~\cite{Dias:2026Companion} (following the strategy in~\cite{Gundlach:1998Angular,Gundlach:2001ey}).
For the tilted ($n\perp\sigma^3$) configuration, a rigid rotation of $n$ about $\sigma^3$, i.e. a change of the phase of $\Phi$, preserves the ansatz and carries angular momentum along $\sigma^3$, so $J$ is linear in the tilt perturbations of BCS. The least-damped tilt mode from our perturbation results give \cite{Dias:2026Companion}
\begin{equation}
\lambda_1=-1.52(1)+{\rm i}\kappa/2\;,\qquad \kappa\equiv2\pi/\Delta\;.
\end{equation}
This particular value of ${\rm Im}\lambda_1$ means that the factor $e^{\lambda_1\Delta}$ by which the perturbation mode is multiplied after one background period is real and negative, and $\lambda_1$ is not part of a complex conjugate pair, as is seen numerically. In other words, the tilt mode
changes sign once per echo, and so does $J_{\rm BH}$.
The corresponding scaling exponent and period are
\be
  \mu=\frac{3-{\rm Re}\,\lambda_1}{\lambda_u}\;,\qquad
  T_J=\frac{2\pi\lambda_u}{{\rm Im}\,\lambda_1}=2T .
  \label{eq:mutilt}
\ee
The numerical value of $\mu$ is shown in Table~\ref{tab:exponents} and agrees with the $c=0$ time evolution data.

For the aligned solution, symmetry excludes this linear contribution. BCS is invariant under rotations about its own axis, so rotating it about $\sigma^3\parallel n$ generates no angular momentum, while perturbations that tilt $n$ away from $\sigma^3$ are excluded by the ansatz. Angular momentum is instead carried by perturbations of $\Phi$, which describe triaxial deformations about $n$, proportional to $\sigma_\pm^2$. Since $\mathcal J'=4r^3(X\Pi_Y-Y\Pi_X)$ is quadratic, $J$ is a product of two modes. This is the vacuum counterpart of the second-order mechanism of Ref.~\cite{Garfinkle:1999Angular}. On the aligned background, $X$ and $Y$ obey the same real linear equation, so nonzero $J$ requires a combination of different $X$ and $Y$ eigenmodes. The relevant pair 
are the two least damped modes that generate non-zero $\Phi$:
\begin{equation}
\lambda_a=-1.77\;,\qquad \lambda_b=-1.85+{\rm i}\kappa/2\;,
\end{equation}
As $e^{(\lambda_a+\lambda_b)\Delta}$ is real and negative, $J_{\rm BH}$ again changes sign once per echo, and its critical exponent is expressed as 
\be
  \mu=\frac{3-{\rm Re}\,\lambda_a-{\rm Re}\,\lambda_b}{\lambda_u},\qquad T_J=2T.
  \label{eq:mualigned}
\ee
Again, the numerical value of $\mu$ is shown in Table~\ref{tab:exponents} and agrees with the $c=1$ time evolution data.


We also tested the perturbative result that $T_J=2T$ with subsequent time evolutions, as this equation predicts when the next sign change occurs.  The relative discrepancy in the location of this sign change is $10^{-3}$.

The fact that the tilted perturbations generate angular momentum at first order, while the aligned perturbations only do so at second order also explains the vast difference in angular momentum seen between the $c=0$ and $c=1$ time evolution data.  At comparable mass, $|J_{\rm BH}|$ from the $c=1$ data is smaller by three to four orders of magnitude.


{\bf Discussion.} Near-threshold black holes in this ansatz are asymptotically non-rotating. More precisely, their dimensionless spin scales as $J_{\rm BH}/M_{\rm BH}^{3/2}\propto\delta^{\mu-3\gamma/2}$, with $\mu-3\gamma/2\simeq0.25$ for the tilted solution and $0.60$ for the aligned one. The initial data carry angular momentum, but near threshold, perturbations of the critical background cause the angular momentum to decay away. 
Away from threshold, the same system admits solutions in which gravitational waves carrying angular momentum form an extremal black hole in finite time~\cite{Crump:2026kgu}.

Within the SU$(2)\times\mathbb Z_4$ symmetry class, the mass exponent $\gamma$ is universal, but the angular momentum exponent $\mu$ is not. The initial data determine which embedding (orientation) of BCS is approached and hence which angular-momentum exponent is observed. Even with $c=0$ (initial data with $B=0$), for instance, the profile $f$ of (\ref{eq:ID}) leads to the tilted configuration, while $f=r^2e^{-r^2/2}$ leads to the aligned one. Perturbation theory about a non-rotating critical solution predicts that $\mu$ is set by the least-damped perturbation carrying $J$ at linear order, or, when symmetry forbids a linear contribution, by a product of perturbations~\cite{Gundlach:1998Angular,Gundlach:2001ey,Garfinkle:1999Angular}. Here both mechanisms occur for different embeddings of the same critical solution. The distinction arises because the vacuum critical solution is  anisotropic: its orientation relative to $J$ is an additional datum that is absent for spherically symmetric critical solutions.

With relaxed symmetry assumptions, we expect the tilted exponent to govern generic collapse. If non-trivial $\sigma^1\sigma^3$ and $\sigma^2\sigma^3$ components are allowed, the angle $\theta$ between $n$ and the rotation axis becomes continuous, and a rotation of $n$ that changes $\theta$ becomes a zero mode. Both mechanisms can then contribute, $J_{\rm BH}\simeq a\sin\theta\,\delta^{0.743}+b\,\delta^{1.088}$, and the first term dominates for sufficiently small $\delta$: $\delta\lesssim(\sin\theta)^{1/0.345}$. Only $\theta=0$ exactly gives $\mu=1.088$, which therefore requires tuning of one further parameter, as the aligned symmetry itself does. Testing this expectation requires evolutions beyond the present ansatz \eqref{eq:ansatz}, though still possible with a $(1+1)$ system.

Several questions remain. Within the ansatz, the boundary between the two basins has codimension-two; a natural candidate for the solution on this boundary is the triaxial codimension-two solution in \cite{Bizon:2006Codim}. 
One may also wonder if there is a structural reason why $e^{\lambda_1\Delta}$, $e^{\lambda_b\Delta}$ are negative, and ${\rm Im}\,\lambda_1={\rm Im}\,\lambda_b=\kappa/2$. 
Finally, to our knowledge, every critical solution found thus far is irrotational, and our results give no indication that a rotating critical solution exists.  It would be interesting to lift the cohomogeneity-one restriction and to study rotating critical collapse in four dimensions, where even the non-rotating vacuum critical solution remains elusive~\cite{Baumgarte2023,BaumgarteGundlachHilditch2026,Adhikari2026}.

{\it Acknowledgments.}
OJCD acknowledges financial support from the  STFC consolidated grant~ST/X000583/1.
JES is supported by STFC grant number ST/X000664/1 and partially by Hughes Hall, Cambridge.

\bibliographystyle{apsrev4-2}
\bibliography{prl}

\clearpage

\onecolumngrid

\begin{center}
  {\large\bf Supplemental Material}
\end{center}

\setcounter{equation}{0}
\setcounter{table}{0}
\setcounter{figure}{0}
\renewcommand{\theequation}{S\arabic{equation}}
\renewcommand{\thetable}{S\arabic{table}}
\renewcommand{\thefigure}{S\arabic{figure}}
\renewcommand{\theHequation}{S\arabic{equation}}
\renewcommand{\theHtable}{S\arabic{table}}
\renewcommand{\theHfigure}{S\arabic{figure}}

\subsection{Left invariant $1$-forms}

The left-invariant $1$-forms on SU$(2)$ are
\begin{equation}
\begin{aligned}
&\sigma^1 = \cos \psi\,\sin \theta\,{\rm d}\phi-\sin \psi\,{\rm d}\theta\,,
\\
&\sigma^2 = \sin \psi\,\sin \theta\, {\rm d}\phi+\cos \psi\,{\rm d}\theta\,,
\\
&\sigma^3={\rm d}\psi+\cos \theta \,{\rm d}\phi\,,
\end{aligned}
\end{equation}%
with $\theta\in[0,\pi]$, $\phi\sim \phi+2\pi$ and $\psi\sim \psi+4\pi$.

\subsection{Field equations}

We write $s\equiv\sqrt{1+|\Phi|^2}$ and, for complex $U=U_X+{\rm i}U_Y$ and $V=V_X+{\rm i}V_Y$,
\be
  G(U,V)\equiv\frac{(1+Y^2)U_XV_X-XY(U_XV_Y+U_YV_X)+(1+X^2)U_YV_Y}{1+|\Phi|^2},
\ee
is the hyperbolic metric on the $\Phi$ plane. The target-space metric of the sigma model for $(B,\Phi)$ is $3\,{\rm d}B^2+G({\rm d}\Phi,{\rm d}\Phi)$, and the curvature of the orbits enters through
\be
  V(B,\Phi)\equiv4e^{-4B}-16e^{-B}s+16e^{2B}|\Phi|^2 ,
\ee
with $V=-12$ on the round three-sphere. In the following, we denote radial derivatives by a prime.

{\it Maximal-areal gauge.} In this gauge, we choose
\be\label{eq:MaxAreal}
  g_{ab}{\rm d}y^a{\rm d}y^b=-\alpha^2{\rm d}t^2+\frac{({\rm d}r+\beta\,{\rm d}t)^2}{A},\qquad
  \mathcal A=\Omega\,{\rm d}t ,
\ee
with maximal slicing, ${\rm tr}\,K=0$. We write $K\equiv K^r{}_r$; since $r$ is areal, the trace of $K$ over the orbits is $3\beta/(r\alpha)=-K$. The momenta of the main text are $\Pi_B=(\partial_tB-\beta B')/(\alpha\sqrt A)$ and similarly for $\Phi$, so
\bea
  \partial_tB&=&\alpha\sqrt A\,\Pi_B+\beta B',\\
  \partial_tX&=&\alpha\sqrt A\,\Pi_X+\beta X'+4\Omega Y,\\
  \partial_tY&=&\alpha\sqrt A\,\Pi_Y+\beta Y'-4\Omega X .
\eea
The remaining fields are determined on each slice. The momentum and Hamiltonian constraints are
\bea
  (r^4K)'&=&-\frac{r^4\sqrt A}{2}\left[3\,\Pi_BB'+G(\Pi_\Phi,\Phi')\right],\\
  A'+A\left[\frac2r+\frac r6\left(3B'^2+G(\Phi',\Phi')\right)\right]
  &=&-\frac{V}{6r}-\frac{e^{2B}\mathcal J^2}{6r^7}
  -\frac{rA}{6}\left[3\Pi_B^2+G(\Pi_\Phi,\Pi_\Phi)\right]-\frac{4r}{9}K^2 ,
\eea
the angular momentum and the gauge field follow from
\be
  \mathcal J'=4r^3\left(X\Pi_Y-Y\Pi_X\right),\qquad
  \Omega'=\frac{\alpha\,e^{2B}\mathcal J}{r^5\sqrt A},\qquad \Omega=0\ \text{at the outer boundary},
\ee
and the lapse and shift from
\be
  \left(r^3\sqrt A\,\alpha'\right)'=\frac{r^3\alpha K_{ij}K^{ij}}{\sqrt A},\qquad
  \beta=-\frac{r\alpha K}{3}.
\ee
The outgoing null expansion of an orbit vanishes where $\sqrt A+rK/3=0$. The evolution equations for the momenta are lengthy and not displayed here.

{\it Polar-areal gauge.} Here
\be \label{eq:PolarAreal}
  g_{ab}{\rm d}y^a{\rm d}y^b=-Ae^{-2\delta}{\rm d}t^2+\frac{{\rm d}r^2}{A},\qquad
  \mathcal A=\Omega\,{\rm d}t,\qquad \delta(t,0)=\Omega(t,0)=0 ,
\ee
so $t$ is central proper time. The fields evolve as
\be
  \partial_tB=Ae^{-\delta}\Pi_B,\qquad
  \partial_tX=Ae^{-\delta}\Pi_X+4\Omega Y,\qquad
  \partial_tY=Ae^{-\delta}\Pi_Y-4\Omega X ,
\ee
and the slice equations are radial quadratures,
\bea
  \delta'&=&-\frac r6\left[3\left(B'^2+\Pi_B^2\right)+G(\Phi',\Phi')+G(\Pi_\Phi,\Pi_\Phi)\right],\\
  \left(r^2Ae^{-\delta}\right)'&=&-r^2e^{-\delta}\left(\frac{V}{6r}+\frac{e^{2B}\mathcal J^2}{6r^7}\right),\\
  \mathcal J&=&r^5e^{-2B+\delta}\,\Omega',\qquad \mathcal J'=4r^3\left(X\Pi_Y-Y\Pi_X\right).
\eea
For $B=\Phi=\mathcal J=0$ both gauges give $A=1-m/r^2$, the five-dimensional Schwarzschild-Tangherlini solution.

{\it Normalisation of $J$.} With $G_5=1$, the angular momentum within the orbit of radius $r$ is $J(r)=\pi\mathcal J(r)/16$, which is the Komar angular momentum wherever $\Phi=0$, and $\mathcal J=r^5e^{-2B}\sqrt A\,\Omega'/\alpha$ in maximal-areal gauge and $\mathcal J=r^5e^{-2B+\delta}\Omega'$ in polar-areal gauge. $J_{\rm BH}$ is this quantity on the apparent horizon.

{\it Initial data.} The data (\ref{eq:ID}) have $X=\Pi_B=\Pi_Y=0$, so the right-hand side of the momentum constraint vanishes and $K=0$, hence $\beta=0$. The initial slice is therefore both maximal and polar, and the two codes evolve the same Cauchy data. The Hamiltonian constraint then fixes $A$, and the data carry $\mathcal J(\infty)=-20\,\varepsilon^2\int_0^\infty r^3f^2\,{\rm d}r<0$. The finite-difference code was derived with the opposite orientation of the one-forms, $\sigma^i\to-\sigma^i$, which together with $\mathcal A\to-\mathcal A$ leaves Eq.~(\ref{eq:ansatz}) invariant; it therefore evolves the mirror image of these data, with $\mathcal J>0$, and thresholds, exponents and periods are unaffected.

\subsection{Equal-angular-momenta Myers–Perry black hole}

The equal-angular-momenta Myers–Perry (MP) black hole is a vacuum solution with isometry group $\mathbb{R}\times$SU$(2)\times$U$(1)$ (i.e. with $\Phi=0$), often written in Boyer–Lindquist coordinates $(\tilde{t},\tilde{r},\theta,\phi,\tilde{\psi})$, as \cite{Myers:1986un,Gibbons:2004uw,Gibbons:2004js,Kunduri:2006qa,Dias:2010eu}
\begin{subequations} \label{MPoriginal}
\begin{align}
& {\rm d}s^2=-\frac{f(\tilde{r})}{h(\tilde{r})}{\rm d}\tilde{t}^2+\frac{{\rm d}\tilde{r}^2}{f(\tilde{r})}+\frac{\tilde{r}^2}{4}\Big[\sigma_+ \sigma_-
+h(\tilde{r})\left(\tilde{\sigma}^3+2 \omega(\tilde{r}){\rm d}\tilde{t} \right)^2\Big] \\
& \hbox{with} \quad h(\tilde{r})=1+\frac{\tilde{r}_0^2 a^2}{\tilde{r}^4}\,,\quad f(\tilde{r})=1+\frac{\tilde{r}_0^2 a^2}{\tilde{r}^4}-\frac{\tilde{r}_0^2}{\tilde{r}^2}\quad\text{and}\quad \omega(\tilde{r})=\frac{a \tilde{r}_0^2}{\tilde{r}^4h(\tilde{r})}\,.
\end{align}
\end{subequations}
Here, $\tilde{r}_0$ and $a$ are, respectively, mass and rotation parameters with $\tilde{r}_0>0$ and $|a|\leq\tilde{r}_0/2$. The event horizon is located at $\tilde{r}=\tilde{r}_+$, with $\tilde{r}_0=\frac{\tilde{r}_+^2}{\sqrt{\tilde{r}_+^2-a^2}}$, which is generated by the Killing vector $\partial_{\tilde{t}}-2\omega(\tilde{r}_+)\partial_{\tilde{\psi}}$. When $|a|=\tilde{r}_0/2$, one has a smooth extremal solution with zero temperature. The angular momenta in the two orthogonal planes of rotation are equal, and the rotation is along $\tilde{\sigma}^3$. The Komar mass and angular momenta of the MP black hole agree with the Arnowitt-Deser-Misner mass and angular momenta and are given by $M_{\rm BH}= \frac{3\pi}{8 G_5}\,\tilde{r}_0^2 $ and $J_{\rm BH}= -\frac{\pi}{4 G_5}\,\tilde{r}_0^2\,a$, respectively, with the sign convention for $J$ of the previous subsection.
 
{\it MP in the maximal-areal gauge.}
 The MP metric takes the maximal-areal form after the change of coordinates
\begin{align}\label{eq:MP-maxAreal1}
  &  \tilde{t} =t+ \int P(r){\rm d}r \quad \hbox{with} \quad P(r)=-\frac{C\sqrt{h(z)}z^{\prime}}{f(z) \sqrt{C^2 +f(z)z^6}}, \nonumber\\  
  & \tilde{r}=z(r) \quad \hbox{with} \quad z(r)=\frac{\sqrt{2^{1/3}\left(9 r^6+\sqrt{12 a^6 \tilde{r}_0^6+81 r^{12}}\right)^{2/3}-2\cdot 3^{1/3} a^2 \tilde{r}_0^2}}{6^{1/3} \left( 9 r^6+ \sqrt{12 a^6 \tilde{r}_0^6+81 r^{12}}\right)^{1/6}}, \nonumber\\
   & \tilde{\sigma}^3=\sigma^3+\varpi(r) {\rm d}r   \quad \hbox{with} \quad \varpi(r)=-2P(r)\omega(z), 
\end{align}
where we use the short notation $Q(z)\equiv Q(z(r))$ and note that the inverse radial map is $r=r(\tilde{r})=\tilde{r}\,h(\tilde{r})^{1/6}$. This makes $r$ the areal radius of the orbits.
It follows that the equal-angular-momenta MP black hole (\ref{MPoriginal}) is described in the maximal-areal gauge by (\ref{eq:ansatz}) and (\ref{eq:MaxAreal}) with
\begin{align}\label{eq:MP-maxAreal2}
  &  B(r)=-\frac{1}{3}\ln h(z), \quad A(r)=\frac{C^2 +f(z)z^6}{z^6 \left(z^{\prime}\right)^2}, \quad \beta(r)=\frac{C \sqrt{A}}{r^3},\quad \alpha=\frac{\sqrt{A} z^{\prime}}{\sqrt{h(z)}},\quad  \Omega(r)=\omega(z), \quad \Phi(r)=0.
\end{align}
In (\ref{eq:MP-maxAreal1})--(\ref{eq:MP-maxAreal2}), $C$ is a constant that follows from requiring that the trace of the extrinsic curvature vanishes, ${\rm tr}\,K=\frac{\sqrt{A}}{\alpha\, r^3}\frac{{\rm d}}{{\rm d}r}\left(\frac{r^3 \beta}{\sqrt{A}}\right)=0\Rightarrow \frac{r^3 \beta}{\sqrt{A}}=C$. The field equations leave $C$ free; it labels the maximal slice and is set by the slicing. Once the exterior of a black hole formed in collapse has settled to MP, $r^3\beta/\sqrt{A}$ is independent of $r$ there and gives $C(t)$, whose value is fixed by the boundary conditions of the lapse equation rather than by the initial data. With a regular centre, $C(t)$ tends at late times to the value at which the slices approach the limit surface, $C^2=\max_{f(z)<0}\left[-f(z)\,z^6\right]$, where $A\to0$ and the lapse collapses; this depends only on $\tilde r_0$ and $a$.

{\it MP in the polar-areal gauge.}
  The MP solution already has no cross term ${\rm d}\tilde{t}{\rm d}\tilde{r}$ so a direct matching of the metric components of (\ref{MPoriginal}) with (\ref{eq:ansatz}) and (\ref{eq:PolarAreal}) yields the map $\tilde{t}=t, \tilde{r}=z(r)$ and 
  \begin{equation}\label{eq:MP-PolarAreal}
   B(r)=-\frac{1}{3}\ln h(z), \qquad A(r)=\frac{f(z)}{\left(z^{\prime}\right)^2}, \qquad \delta(r)=\frac{1}{2}\ln \left( \frac{h(z)}{\left(z^{\prime}\right)^2} \right),\qquad   \Omega(r)=\omega(z), \qquad \Phi(r)=0,
\end{equation}
where $z(r)$ is defined in (\ref{eq:MP-maxAreal1}), i.e. $r=\tilde{r}\,h(\tilde{r})^{1/6}$. Since the MP exterior has no regular centre, $\delta$ and $\Omega$ are normalised here to vanish at infinity rather than at $r=0$ as in Eq.~(\ref{eq:PolarAreal}).

\subsection{Numerical methods}

{\it Discontinuous-Galerkin code (maximal-areal gauge).} The fields are represented on a nodal Legendre-Gauss-Lobatto discontinuous-Galerkin grid in the areal radius, $r\in[0,40]$ and a variant of this code was already used in the AdS gravitational collapse study of \cite{Choptuik:2017cyd}. The initial grid has 100 elements, most of which have polynomial order $p=16$ ($p=20$ in the innermost four). Every 50 steps the grid is adapted in $h$ and $p$ from the decay of the Legendre coefficients in each element, with a truncation target of $10^{-8}$; elements requiring an order above $p=24$ are split. Time integration is fourth-order Runge-Kutta at Courant number 1, with an exponential filter and local Lax-Friedrichs fluxes. The slice equations (constraints, $\Omega$, lapse and shift) are solved on every stage by block-bidiagonal marching, and the initial data by Newton iteration to $10^{-14}$. Once an apparent horizon has formed, the region with outgoing expansion below $-0.05$ is excised, and the outer boundary at $r=40$ is outgoing. The Mathematica and Julia implementations agree to $10^{-12}$. $J_{\rm BH}$ and the horizon radius are averaged over a late-time window, $t\in[13,14]$ at $c=0$ and $t\in[12,13]$ at $c=1$. Thresholds are found by asynchronous multisection on six workers.

{\it Finite-difference code (polar-areal gauge).} The slice equations are solved as fourth-order cumulative radial quadratures on every Runge-Kutta stage, so the evolution is fully constrained. Derivatives are fourth-order centred differences on a vertex-centred grid with parity ghosts at $r=0$, with sixth-difference Kreiss-Oliger dissipation, and time integration is fourth-order Runge-Kutta at Courant number $0.25$. Mesh refinement uses up to 40 levels, each halving the spacing and nested at the origin, with a common time step. The levels are regridded every 8 steps from undivided second differences of the scale-invariant fields. The outer boundary is at $r=20$, with a sponge on its outer quarter. The base spacing is $h_0=0.02$, and $h_0=0.01$ for the resolution test. In polar-areal gauge, horizon formation creates a coordinate singularity, which causes numerical breakdown of the code.  Collapse is declared when $\max(1-A)>0.9$, which stops the code just before horizon formation.  The horizon radius and $J_{\rm BH}=\pi\mathcal J/16$ are read at the minimum of $A$. This proxy provides sufficient accuracy for $J$, but not for the mass: along a finite-difference sequence the residuals of $\ln M$ about a fit of Eq.~(\ref{eq:Mfit}) have an rms of about $0.1$, against $4\times10^{-4}$ in the discontinuous-Galerkin code, which is too large to resolve the modulation $\psi$. 
We therefore use this code only to measure thresholds, orientation, $\mu$ and $T_J$. Bisection is carried to the double-precision limit.

\subsection{Fits}

The mass is fitted by Eq.~(\ref{eq:Mfit}) with
$\psi(L)=\sum_{k=1}^{n_h}\left[a_k\cos(2\pi kL/T)+b_k\sin(2\pi kL/T)\right]$ and $L=\ln\delta$, taking $n_h=3$ at $c=0$ and $n_h=2$ for the shorter $c=1$ sequence (Table~\ref{tab:windows}). At $c=0$ the second and third harmonics have amplitudes of $13\%$ and $2\%$ of the fundamental.
 The angular momentum is fitted by Eq.~(\ref{eq:Jfit}) with residuals weighted by the envelope, $[J_{\rm BH}-A_J\delta^\mu\cos(\cdots)]/(A_J\delta^\mu)$, since $|J_{\rm BH}|$ varies by a factor of order $10^3$ along a sequence; an unweighted fit is dominated by the least fine-tuned points (Table~\ref{tab:Jfits}). Errors are the larger of the covariance and jackknife estimates, and are statistical only. The value of $\gamma$ has some dependence on the window in $L$, because corrections to scaling affect the least fine-tuned points, while $T$ does not (Table~\ref{tab:windows}); we quote the most fine-tuned window. In the finite-difference sequences, points with $L<-32$ are discarded: there $\delta$ is comparable to the width of the final bisection bracket, and $L$ is not reliable.

\begin{table}[htbp]
  \caption{Window dependence of the mass fit, DG, $c=0$.}
  \label{tab:windows}
  \begin{ruledtabular}
  \begin{tabular}{lccc}
    window in $L$ & points & $\gamma$ & rms of $\ln M$ \\
    \hline
    $[-18.4,-14.3]$ & 18 & 0.32885(6)  & $5.8\times10^{-5}$ \\
    $[-14.4,-10.0]$ & 22 & 0.32985(46) & $1.7\times10^{-4}$ \\
    all             & 39 & 0.32911(7)  & $3.7\times10^{-4}$
  \end{tabular}
  \end{ruledtabular}
\end{table}

\begin{table}[htbp]
  \caption{Angular-momentum fits.}
  \label{tab:Jfits}
  \begin{ruledtabular}
  \begin{tabular}{llcccc}
    code & family & window in $L$ & points & $\mu$ & $T_J$ \\
    \hline
    DG & $c=0,$ $f$ in \eqref{eq:ID} & all & 39 & 0.737(9) & 5.658(46) \\
    FD, $h_0=0.02$ & $c=0,$ $f$ in \eqref{eq:ID}& $(-32,-9)$ & 61 & 0.745(5) & 5.721(25) \\
    FD, $h_0=0.01$ & $c=0,$ $f$ in \eqref{eq:ID}& $(-32,-9)$ & 56 & 0.740(7) & 5.718(20) \\
    DG & $c=1,$ $f$ in \eqref{eq:ID}& all & 16 & 1.089(25) & 5.375(93) \\
    DG & $c=1,$ $f$ in \eqref{eq:ID}& $<-12$ & 12 & 1.077(29) & 5.624(116) \\
    FD, $h_0=0.02$ & $c=1,$ $f$ in \eqref{eq:ID}& $(-32,-9)$ & 36 & 1.088(6) & 5.608(67) \\
    FD, $h_0=0.02$ & $c=0$, $f\to r^2e^{-r^2/2}$ & $(-32,-9)$ & 64 & 1.084(4) & 5.688(25)
  \end{tabular}
  \end{ruledtabular}
\end{table}

\subsection{Convergence and cross-code tests}

The two codes find the same threshold for identical data (Table~\ref{tab:thresholds}). The finite-difference threshold changes by $2\times10^{-11}$ between $h_0=0.02$ and $0.01$, and is converged. The discontinuous-Galerkin threshold moves toward it as the truncation target is lowered from $10^{-8}$ to $10^{-9}$, so the remaining difference is the discretisation error of that code.

\begin{table}[htbp]
  \caption{Threshold amplitudes $\varepsilon_\star$ of the family (\ref{eq:ID}).}
  \label{tab:thresholds}
  \begin{ruledtabular}
  \begin{tabular}{lcc}
    & $c=0$ & $c=1$ \\
    \hline
    DG, truncation $10^{-8}$ & 3.4173191791 & 1.31090144907 \\
    DG, truncation $10^{-9}$ & 3.4173189385 & -- \\
    FD, $h_0=0.02$ & 3.41731819397 & 1.31090143859 \\
    FD, $h_0=0.01$ & 3.41731819404 & -- \\
    relative difference, DG ($10^{-8}$) $-$ FD & $3\times10^{-7}$ & $8\times10^{-9}$
  \end{tabular}
  \end{ruledtabular}
\end{table}

At truncation $10^{-9}$ we repeated eight members of the $c=0$ sequence, with $\delta$ between $3\times10^{-7}$ and $3\times10^{-5}$, measured from the new threshold. Holding the fine structure $\psi$ fixed, we find that the new masses agree with the $10^{-8}$ curve, with no shift along it ($0.002\pm0.005$ in $L$) and a change of slope $\Delta\gamma=(0.8\pm1.1)\times10^{-4}$ in the fine-tuned window ($(1.0\pm1.1)\times10^{-3}$ over all usable points). $J_{\rm BH}$ at equal $L$ agrees with the $10^{-8}$ fit within the scatter of that fit, about $9\%$ of the envelope, and the zero crossings coincide. In the finite-difference code, $J_{\rm BH}$ at equal $L$ agrees between $h_0=0.02$ and $0.01$ to a few per cent.

As a test of the $\Phi$ sector, we rotate BCS data into the $\sigma^1$-$\sigma^2$ plane, so that the non-rotating BCS problem is posed in rotated variables. The finite-difference evolution should preserve this orientation and reproduce the non-rotating threshold. We find that the thresholds agree to $8\times10^{-8}$, the orientation measure $\vartheta$ defined below remains at $1.000$, and the last echoes give $\Delta=0.4671$, consistent with the value $0.46717$ obtained from the self-similar solution.

\subsection{Orientation}

In the discontinuous-Galerkin evolutions we record the fields at fixed radii during the echoing phase of a near-critical member of each family (at $c=1$ the most fine-tuned supercritical run of the bisection, $\delta\sim3\times10^{-10}$) and compare $\operatorname{arcsinh}|\Phi|$ with $3|B|$ (Table~\ref{tab:orient}). In the finite-difference code we use $\vartheta\equiv\operatorname{arcsinh}|\Phi|/(3|B|)$ at the centre, which is $0$ for the aligned and $1$ for the tilted orientation; it is $1.000$ at $c=0$ and $0.000$ at $c=1$. At $c=0$ and $r=0.1$, $\operatorname{arcsinh}|\Phi|/|B|$ lies in the range $2.98$-$3.01$ over the last two echoes while $|B|$ reaches $0.35$; the linear ratio $|\Phi|/|B|$ exceeds 3 there only because of the nonlinearity of $\operatorname{arcsinh}$. At $c=0$ and $r=0.01$ the two nearly equal stretches agree to about $2\times10^{-4}$ throughout, while the anisotropy $(s_3-s_1)/s_2$ reaches $0.63$; at $r=0.1$ their relative difference falls from echo to echo, from $4\times10^{-2}$ to $7\times10^{-3}$. At $c=1$ the stretches $e^{B}(\sqrt{1+|\Phi|^2}\pm|\Phi|)$ differ by a fraction $\simeq2|\Phi|$.

\begin{table}[htbp]
  \caption{Orientation diagnostics near the centre, DG.}
  \label{tab:orient}
  \begin{ruledtabular}
  \begin{tabular}{lccc}
    & anisotropy at $c=1$, up to & $\operatorname{arcsinh}|\Phi|/|B|$, $c=0$ & $|\Phi|/|B|$, $c=1$ \\
    \hline
    $r=0.01$, early echoes & 0.09 & -- & 0.002 \\
    $r=0.01$, late echoes & 0.88 & 3.01 & 0.001 \\
    $r=0.1$, successive echoes & 0.6-2.2 & 2.98-3.01 & 0.056, 0.020, 0.009, 0.005
  \end{tabular}
  \end{ruledtabular}
\end{table}

\subsection{Basins of the two orientations}

With the finite-difference code we bisected the family (\ref{eq:ID}) for several $c$, to about $10^{-8}$, which is sufficient to identify the orientation but not to measure $\mu$ (Table~\ref{tab:basins}). The boundary between the basins lies at $0.3<|c|<1$. For $c=-3$ we found no threshold for $\varepsilon\in[0.5,6]$. The family with $f$ replaced by $r^2e^{-r^2/2}$ at $c=0$ reaches the aligned orientation, so the orientation depends on the radial profile as well as on $c$.

\begin{table}[htbp]
  \caption{Orientation of the critical solution across the family (\ref{eq:ID}), FD.}
  \label{tab:basins}
  \begin{ruledtabular}
  \begin{tabular}{lcc}
    $c$ & $\vartheta$ & orientation \\
    \hline
    $0$, $\pm0.3$ & 1.00 & tilted \\
    $\pm1$, $3$ & 0.00 & aligned \\
    $-3$ & -- & no threshold in $\varepsilon\in[0.5,6]$
  \end{tabular}
  \end{ruledtabular}
\end{table}

\subsection{The critical solution and its perturbations}

The BCS solution is computed as an exactly periodic solution in self-similar variables by Newton iteration, seeded by near-critical data from a separate non-rotating evolution bisected to $10^{-14}$, with convergence to machine precision at fixed resolution and convergence under grid refinement, giving $\kappa=13.44945325$ and $\Delta=2\pi/\kappa=0.46717031$. Its Floquet spectrum in the biaxial sector contains one growing mode, apart from the gauge modes $\lambda=1$ (a shift of the accumulation time) and $\lambda=0$ (a shift of the echoing phase). The next biaxial mode, $\lambda=-0.383\pm1.061{\rm i}$, carries no angular momentum. The tilt and triaxial ($\sigma_\pm^2$) sectors are derived from the five-dimensional Ricci tensor and solved on the same background. A static rigid rotation of $n$ about $\sigma^3$, an exact zero mode, is recovered at $\lambda=0.007$, which provides an estimate of the numerical error. Details are given in Ref.~\cite{Dias:2026Companion}.

\begin{table}[htbp]
  \caption{The growing mode of the BCS solution on successive grids. $T$ uses the echoing period $\Delta$ of the background computed on the same grid.}
  \label{tab:floquet}
  \begin{ruledtabular}
  \begin{tabular}{lccc}
    grid & $\lambda_u$ & $\gamma=2/\lambda_u$ & $T=\lambda_u\Delta$ \\
    \hline
    $49\times57$ & 6.0910 & 0.32835 & 2.8457 \\
    $57\times65$ & 6.0848 & 0.32869 & 2.8426 \\
    $57\times65$ (larger domain) & 6.0865 & 0.32860 & 2.8432 \\
    $65\times73$ & 6.0861 & 0.32862 & 2.8432
  \end{tabular}
  \end{ruledtabular}
\end{table}

\subsection{Quasinormal ringdown}

As a test of both the evolution code and the rotating sector of the reduced equations, we compare the ringdown of the black holes formed in our evolutions with the quasinormal modes (QNMs) of the equal-angular-momenta Myers-Perry black hole, computed within the same symmetry class.

{\it Background.} In terms of the radius $\rho$ of Ref.~\cite{Kunduri:2006qa}, with horizon at $\rho_+$, mass parameter $\tilde{\mu}$ and rotation parameter $a$, the Myers-Perry solution in the polar-areal variables above is
\begin{multline}
  r=\rho\,q^{1/6},\qquad e^{B}=q^{-1/3},\qquad \Omega=\frac{\tilde{\mu} a}{\rho^4 q},\qquad
  \mathcal J=-4\tilde{\mu} a,
  \\
  Ae^{-2\delta}=\frac{1}{q}\left(1-\frac{\tilde{\mu}}{\rho^2}+\frac{\tilde{\mu} a^2}{\rho^4}\right),\qquad
  A=\left(\frac{dr}{d\rho}\right)^2\left(1-\frac{\tilde{\mu}}{\rho^2}+\frac{\tilde{\mu} a^2}{\rho^4}\right),
\end{multline}
with $q\equiv1+\tilde{\mu} a^2/\rho^4$, $\Phi=0$, and $\Omega,\delta\to0$ at infinity (for a black-hole background we normalise at infinity rather than at $r=0$ as in the evolutions; the two normalisations differ by a constant rescaling of $t$ and a constant shift of $\Omega$, so that frequencies refer to static observers at infinity and $\omega-4\Omega$ is unchanged); for $a>0$ it has $\mathcal J<0$, the sign of $J$ in our evolutions. This solves the slice equations and the static $B$ equation. Its spin is $\chi\equiv27\pi J^2/(8M^3)=4\hat a^2(1-\hat a^2)$, with $\hat a=a/\rho_+$, so that $\chi=1$ at extremality.

{\it Linear perturbations.} At linear order the two sectors decouple. Because $\Phi$ enters every constraint quadratically, it obeys a single equation on the fixed background, with charge four under $\mathcal A$:
\begin{equation}
  Ae^{-\delta}\left(\Phi''+\frac{\Phi'}{r}\right)
  +\frac{e^{-\delta}}{r}\left(\frac{8e^{-B}-2e^{-4B}}{3}-\frac{e^{2B}\mathcal J^2}{6r^6}\right)\Phi'
  +\frac{e^{-\delta}}{r^2}\left(8e^{-B}-16e^{2B}\right)\Phi
  +\frac{e^{\delta}}{A}\left(\omega-4\Omega\right)^2\Phi=0,
  \label{eq:phiqnm}
\end{equation}
for perturbations $\propto e^{-{\rm i}\omega t}$. Denoting linear perturbations by a subscript 1, the perturbation $B_1$ is neutral. The background has $\Pi_B=0$ and solves the static $B$ equation, so $\delta_1$ drops out of the linearised $\Pi_B$ equation. The momentum constraint gives $A_1=-rAB'B_1$, and we have checked that this is consistent with the linearised Hamiltonian constraint. Since $\mathcal J_1=0$, $\Omega$ does not enter, and $B_1$ also obeys a single equation. We solve both equations as quadratic eigenvalue problems by Chebyshev collocation~\cite{Dias:2015nua}, with ingoing behaviour at the horizon, $(\rho-\rho_+)^{-{\rm i}(\omega-4\Omega_H)/(2\kappa_H)}$ for $\Phi$ and $(\rho-\rho_+)^{-{\rm i}\omega/(2\kappa_H)}$ for $B$, where $\Omega_H\equiv\Omega(\rho_+)$ and $\kappa_H$ are the angular velocity and surface gravity of the horizon, and outgoing behaviour $\rho^{-3/2}e^{{\rm i}\omega\rho}$ at infinity. The eigenvalues are converged to better than $10^{-9}$.

We checked the linear spectra in four ways. For $J=0$, $B$ and $\Phi$ are degenerate, as required by the SU$(2)_R$ symmetry that rotates them into each other, and both reduce to the $\ell=2$ mode of the Schwarzschild-Tangherlini black hole, $\omega r_+=1.510567-0.357537{\rm i}$. With rotation, $B$ stays symmetric under $\omega\to-\bar\omega$, while $\Phi$ splits into co- and counter-rotating modes, and near extremality the co-rotating modes become long-lived, $\omega\to4\Omega_H-{\rm i}\kappa_H(j+\tfrac12)$ with overtone number $j$. A time-domain evolution of the same linear equations reproduces the spectra. Finally, an independent derivation of the $B$ sector from the five-dimensional Ricci tensor, in a different gauge, gives a master equation whose spectrum agrees with ours to all quoted digits.
A selection of fundamental modes is given in Table~\ref{tab:qnm}.

\begin{table}[htbp]
  \caption{Fundamental QNMs of the equal-angular-momenta Myers-Perry black hole in the $B$ and $\Phi$ sectors, $\omega r_+$, with $r_+$ the areal radius of the horizon.}
  \label{tab:qnm}
  \begin{ruledtabular}
  \begin{tabular}{cccc}
    $\chi$ & $\Phi$, co-rotating & $\Phi$, counter-rotating & $B$ \\
    \hline
    0     & $1.5106-0.3575{\rm i}$ & $-1.5106-0.3575{\rm i}$ & $\pm1.5106-0.3575{\rm i}$ \\
    0.328 & $1.9093-0.3325{\rm i}$ & $-1.2449-0.3417{\rm i}$ & $\pm1.5143-0.3368{\rm i}$ \\
    0.750 & $2.3484-0.2567{\rm i}$ & $-1.0946-0.3165{\rm i}$ & $\pm1.4988-0.2932{\rm i}$ \\
    0.922 & $2.6754-0.1722{\rm i}$ & $-1.0191-0.2989{\rm i}$ & $\pm1.4604-0.2614{\rm i}$ \\
    0.976 & $2.8839-0.1065{\rm i}$ & $-0.9795-0.2885{\rm i}$ & $\pm1.4242-0.2456{\rm i}$
  \end{tabular}
  \end{ruledtabular}
\end{table}

{\it Ringdown in the evolutions.} We evolved three supercritical members of the family (\ref{eq:ID}) at $c=0$ with the discontinuous-Galerkin code past horizon formation, recording $B$ and $\Phi$ at fixed areal radii between $2r_+$ and $43r_+$. The first is the least fine-tuned point of the ladder, $\delta=4\times10^{-5}$, which forms a nearly non-rotating hole; the other two, $\varepsilon=3.8$ and $4.0$, are far above threshold and form holes with substantial spin. The final hole is identified from its apparent-horizon radius and $J_{\rm BH}$, as a Myers-Perry black hole with mass (\ref{eq:MMP}).

During the ringdown the maximal slices are still relaxing: the lapse at fixed radius keeps growing slowly towards its stationary value. Coordinate time therefore runs faster than the Killing time of the final hole, and frequencies measured in $t$ come out $1$-$3\%$ low, by an amount that depends on radius. We therefore resample each signal in Killing time, $dt_K=\alpha\,dt/N(r)$, where $N$ is the stationary lapse of the final Myers-Perry hole at that radius and the shift is neglected. We then extract complex frequencies with a matrix-pencil (Prony) fit, as in Ref.~\cite{Choptuik:2017cyd}, in windows that start a light-crossing time after horizon formation.

The results are in Table~\ref{tab:ringdown}. For the $\Phi$ sector they are independent of the extraction radius and of the fit window within the quoted digits, and agree with the linear QNMs to better than $5\times10^{-4}$ in relative terms. This includes the Zeeman splitting of the co- and counter-rotating modes caused by the small residual spin, $\chi=5.6\times10^{-4}$, of the near-threshold hole. In the two spinning cases only the co-rotating $\Phi$ mode is excited, since the initial phase current of $\Phi$ has a definite sign. The neutral $B$ sector agrees to $2\times10^{-4}$ for the near-threshold hole, and to within $0.7\%$ in ${\rm Re}\,\omega$ and $3\%$ in ${\rm Im}\,\omega$ for the spinning ones.

\begin{table}[htbp]
  \caption{Ringdown frequencies, $\omega r_+$, measured in the discontinuous-Galerkin evolutions of the family (\ref{eq:ID}) at $c=0$, against the linear QNMs of the final Myers-Perry black hole. For $B$ the spread over extraction radii and fit windows is shown.}
  \label{tab:ringdown}
  \begin{ruledtabular}
  \begin{tabular}{lccccc}
    $\varepsilon$ & $r_+$ & $\chi$ & mode & evolution & linear \\
    \hline
    $\varepsilon_\star+4\times10^{-5}$ & 0.1391 & $5.6\times10^{-4}$
      & $B$ & $\pm1.5103-0.3575{\rm i}$ & $\pm1.5106-0.3575{\rm i}$ \\
    & & & $\Phi$, co & $1.5230-0.3576{\rm i}$ & $1.5229-0.3575{\rm i}$ \\
    & & & $\Phi$, counter & $-1.4980-0.3575{\rm i}$ & $-1.4984-0.3575{\rm i}$ \\
    \hline
    3.8 & 1.199 & 0.350 & $B$ & $\pm(1.509\textrm{-}1.525)-(0.335\textrm{-}0.343){\rm i}$ & $\pm1.5143-0.3351{\rm i}$ \\
    & & & $\Phi$, co & $1.928-0.330{\rm i}$ & $1.9287-0.3301{\rm i}$ \\
    \hline
    4.0 & 1.404 & 0.707 & $B$ & $\pm(1.497\textrm{-}1.513)-(0.291\textrm{-}0.305){\rm i}$ & $\pm1.5031-0.2992{\rm i}$ \\
    & & & $\Phi$, co & $2.2899-0.2692{\rm i}$ & $2.2907-0.2692{\rm i}$
  \end{tabular}
  \end{ruledtabular}
\end{table}

These initial data carry more angular momentum than a Myers-Perry black hole of the same mass can hold. On the initial slice $\chi=1.51$ at threshold and $1.10$ at $\varepsilon=4$, and the slice is already trapped for $\varepsilon\gtrsim4.2$. Close to threshold almost all of the angular momentum is radiated away. Farther above threshold the hole retains a growing fraction of it, reaching $\chi=0.71$ at $\varepsilon=4$.

\twocolumngrid

\end{document}